\documentclass[%
  aps,preprint,prc,amsmath,amssymb,nofootinbib,superscriptaddress%
]{revtex4}

\usepackage[utf8]{inputenc}
\usepackage[usenames]{color}
\usepackage[dvipsnames]{xcolor}
\usepackage{graphicx}
\usepackage{amsmath}
\usepackage{amsfonts}
\usepackage{amssymb}
\usepackage{mathrsfs}
\usepackage{bm}
\usepackage{verbatim}
\usepackage{cancel}
\usepackage{comment}
\usepackage[normalem]{ulem}
\usepackage{float}
\usepackage{textcomp}
\usepackage{booktabs}

\usepackage{hyperref}
\hypersetup{
  colorlinks=true,        % false: boxed links; true: colored links
  linkcolor=ForestGreen,    % color of internal links
}

\usepackage{multirow,makecell}

\newcommand{\mlo}{M_{\text{lo}}}
\newcommand{\mhi}{M_{\text{hi}}}

\newcommand{\chn}[3]{{{}^{#1}\!{#2}_{#3}}}
\newcommand{\cs}[2]{\chn{#1}{S}{#2}}
\newcommand{\cp}[2]{\chn{#1}{P}{#2}}
\newcommand{\cd}[2]{\chn{#1}{D}{#2}}
\newcommand{\cf}[2]{\chn{#1}{F}{#2}}
\newcommand{\cg}[2]{{}^{#1}{G}_{#2}}
\newcommand{\csd}{{\cs{3}{1}-\cd{3}{1}}}
\newcommand{\cpf}{{\cp{3}{2}-\cf{3}{2}}}
\newcommand{\cdg}{{\cd{3}{3}-\cg{3}{3}}}

\newcommand{\NNLO}{N$^2$LO}

\graphicspath{{./figs/}}

\begin{document}

% \begin{CJK*}{UTF8}{gbsn}

\title{Constructing chiral effective field theory around unnatural leading-order interactions}

% \author{Rui Peng (彭锐)}
\author{Rui Peng}
\affiliation{College of Physics, Sichuan University, Chengdu, Sichuan 610065, China}

% \author{Songlin Lyu (吕松林)}
\author{Songlin Lyu}
\email{songlin@scu.edu.cn}
\affiliation{College of Physics, Sichuan University, Chengdu, Sichuan 610065, China}

\author{Sebastian König}
\affiliation{Department of Phyiscs, North Carolina State University, Raleigh, North Carolina 27695, USA}

% \author{Bingwei Long (龙炳蔚)}
\author{Bingwei Long}
\email{bingwei@scu.edu.cn}
\affiliation{College of Physics, Sichuan University, Chengdu, Sichuan 610065, China}

\date{February 21, 2022}

\begin{abstract}

A momentum-dependent formulation based on a stationary spin-0 and isospin-1 dibaryon field is proposed to improve convergence of chiral effective field theory in the $\cs{1}{0}$ channel of $NN$ scattering. Although the two-parameter leading-order interaction appears to be unnatural, it nevertheless has the necessary features of an effective field theory. A rapid order-by-order convergence is found in $\cs{1}{0}$. As an application beyond the two-body level, the triton binding energy
is studied and compared to standard chiral effective field theory with partly perturbative pions. The consistency of the chiral Lagrangian for the new formulation is examined by working out the pionic radiative corrections, and consequences of nontrivial chiral-connection terms are discussed.
\end{abstract}

\maketitle

% \end{CJK*}

\section{Introduction\label{sec:intro}}

It is a lore about the nuclear force that it has attractive long-range and repulsive short-range components. This can be inferred phenomenologically from the $\cs{1}{0}$ and $\cs{3}{1}$ phase shifts, both showing strong attraction near threshold and vanishing around center-of-mass (c.m.) momentum $k \sim 350$ MeV. But nuclear forces derived from chiral effective field theory (EFT) are characteristically different in the two channels at leading order (LO): in $\csd$, the tensor part of one-pion exchange (OPE) is singularly attractive, providing sufficiently strong attraction at long distances to generate the shallow deuteron bound state. A contact term in $\cs{3}{1}$ provides repulsion at short distances to avoid a collapse of the two-nucleon system into an infinitely deep bound state. This mechanism is well understood to ensure renormalization~\cite{Nogga:2005hy,PavonValderrama:2005gu}. In $\cs{1}{0}$, however, OPE turns out to be numerically weak, thus it can in principle be treated as a perturbation~\cite{Kaplan:1998tg,Kaplan:1998we,Birse:2010jr}. This implies that in $\cs{1}{0}$ the short-range interaction alone needs to provide strong attraction at long distances and strong repulsion at short distances in order to produce the phenomenological behavior of the phase shift. A single $\cs{1}{0}$ contact term at LO as proposed by Weinberg~\cite{Weinberg:1990rz,Weinberg:1991um} is too simple to generate both features simultaneously, which is likely the reason why chiral forces have been found to converge rather slowly in $\cs{1}{0}$~\cite{Long:2012ve,Valderrama:2009ei,Epelbaum:2020maf,Ren:2017yvw}, unless a soft cutoff is carefully chosen~\cite{Valderrama:2009ei}.

As first calculated in Ref.~\cite{Steele:1998zc} and then argued further in Ref.~\cite{Long:2013cya}, the slow convergence can be attributed to the quite large value of the generalized effective range in $\cs{1}{0}$, which is defined in an expansion similar to the effective-range expansion while incorporating iteration of OPE. To improve the convergence, a power counting based on a spin-0 and isospin-1 auxiliary dibaryon field~\cite{Kaplan:1996nv,Soto:2007pg,Soto:2009xy} was suggested in Ref.~\cite{Long:2013cya}, and subsequently developed in Ref.~\cite{SanchezSanchez:2017tws}.

A mere reproduction of empirical phase shifts does not bring much to the discussion of nuclear forces, as there exist several phenomenological models that fit nucleon-nucleon scattering data very well~\cite{Wiringa:1994wb,Stoks:1994wp,Machleidt:2000ge}. We therefore use the following requirements as guidelines to reformulate power counting of chiral forces in order to address the slow $\cs{1}{0}$ convergence. First, as any other consistent EFT, the formulation sets up a controlled expansion, allowing for a highly unnatural LO that includes more than one low-energy constant (LEC) (two in Ref.~\cite{Long:2013cya} and three in Ref.~\cite{SanchezSanchez:2017tws}) in the $\cs{1}{0}$ channel. Second, renormalization-group (RG) invariance is enforced at each order. Third, the symmetries of the chiral Lagrangian are observed, in particular the nonlinearly realized chiral symmetry. The dibaryon formalism has been shown to meet all these requirements, and in particular it does not alter the well-established hierarchy of long-range chiral forces that follows from naive dimensional analysis (NDA)~\cite{Weinberg:1990rz,Weinberg:1991um}: OPE is LO, two-pion-exchange (TPE) contributions are two powers down or more, and so on. This is in contrast to a recent proposal that TPE should be promoted to LO in order to address the convergence issue in $\cs{1}{0}$~\cite{Mishra:2021luw}; see also Ref.~\cite{PavonValderrama:2019lsu} for a related discussion regarding the promotion of TPE.

Unfortunately, the resulting $\cs{1}{0}$ potential is energy-dependent, making it difficult to apply the overall interaction in many-body calculations. For example, the dibaryon-exchange potential, part of LO, is given by
\begin{equation}
  V_\phi(E) = \sigma \frac{y^2}{E + \Delta} \, ,
  \label{eqn_Edep1s0}
\end{equation}
where $E$ is the center-of-mass (c.m.) energy, $\Delta$ the dibaryon mass, $y$ the dibaryon coupling constant, and $\sigma = \pm 1$. In order to facilitate many-body calculations, we discuss in Sec.~\ref{sec:sprpot} a separable, momentum-dependent force to replace the original dibaryon exchange potential in $\cs{1}{0}$
\begin{equation}
  V_\text{spr} (p^\prime, p) = -\frac{4\pi}{m_N} \frac{\lambda}{\sqrt{p^{\prime 2} + m_N \Delta}\sqrt{p^2 + m_N \Delta}}\, ,
  \label{eqn_pdep1s0}
\end{equation}
where $p$ ($p^\prime$) is the incoming (outgoing) c.m. momentum.
The potential~\eqref{eqn_pdep1s0} is clearly motivated by the energy-dependent dibaryon potential~\eqref{eqn_Edep1s0}, and the two potentials~\eqref{eqn_Edep1s0} and~\eqref{eqn_pdep1s0} have the same functional dependence on the energy
when the Born approximation is taken and the nucleons are on the energy shell, $E = {p'}^2/m_N ={p}^2/m_N$ (while still featuring different coupling constants).

Potentials of the same (or similar) form as in Eq.~\eqref{eqn_pdep1s0} were investigated elsewhere in the literature. Reference~\cite{Beane:1997pk} used it to construct a model potential in the context of discussing pionless EFT. More recently, Ref.~\cite{Beane:2021dab} derived a potential with similar square-root form factors based on the so-called UV/IR symmetry of the $S$ matrix, assuming a separable form as in Eq.~\eqref{eqn_pdep1s0}. An attempt to derive a similar potential from the Low equation can be found in Ref.~\cite{Li:2021cue}.

Radiative corrections to nucleon-nucleon contact terms generated by pions (see Fig.~\ref{fig:radpi_diagram}) appear suppressed by $Q^2/\mhi^2$ relative to OPE in a given partial wave. If the contact vertex has a single parameter, e.g., a constant in $S$ waves, the radiative corrections can be shown to generate only terms which are polynomials in momenta or energies because the diagrams are dominated by static intermediate nucleons. Owing to this lack of nonanalytic structures, radiative pions are typically not included in chiral nuclear forces unless chiral extrapolations~\cite{Epelbaum:2002gb,Beane:2002xf,Bai:2021uim} or pion productions are concerned. This is still the case with the energy-dependent dibaryon-exchange. However, as will be discussed in Sec.~\ref{sec:Radpi}, the pionic radiative corrections to the momentum-dependent potential~\eqref{eqn_pdep1s0} can no longer be trivially cast into pure contact form.

In Sec.~\ref{sec:results} we discuss two- and three-body observables calculated with the momentum-dependent potential (plus additional nucleon-nucleon channels) and compare our results to other calculations. As a matter of fact, several nuclear-structure calculations have already used potentials with a similar form to various orders~\cite{Yang:2020pgi,Yang:2021vxa,SanchezSanchez:2020kbx}. The results in Ref.~\cite{Yang:2020pgi} suggest that some nuclei are not bound with the proposed potentials up to NLO, based on which the authors argue that some three-body forces need to be considered at LO. While the role of three-body forces may continue to be debated, we believe that efforts to improve the overall convergence based on two-body and three-body forces can complement each other.

The momentum-dependent dibaryon potential can alternatively be thought of as a way to partially resum derivative-coupled $NNNN$ operators in the effective Lagrangian. With the nonlinearly realized chiral symmetry, derivatives acting on baryon fields must be ``chiral-covariant'' so that chiral symmetry and its spontaneous breaking are properly implemented.
This results in, among other things, a nontrivial $\pi\pi NNNN$ vertex function. We discuss the construction of this vertex and its phenomenological impacts in Sec.~\ref{sec:chiral}. Finally, we summarize and conclude in Sec.~\ref{sec:Summary}.

\section{Momentum-dependent dibaryon potential
\label{sec:sprpot}}

We begin with the $\cs{1}{0}$ power counting developed in Ref.~\cite{Long:2012ve}. Thus, OPE is the LO long-range force, and its $\cs{1}{0}$ projection reads
\begin{equation}
  V_{1\pi}(p^\prime, p) = \frac{g_A^2}{4f_\pi^2}\left( 1 -  \frac{m_\pi^2}{q^2 + m_\pi^2} \right) \, ,
\label{eqn:OPE}
\end{equation}
where the momentum transfer $\vec{q} = \vec{p}\,^\prime - \vec{p}$, the pion mass $m_\pi = 138.0$ MeV, the pion decay constant $f_\pi = 92.4$ MeV, and the axial coupling $g_A = 1.29$. Counting external momenta $Q$, $m_\pi$, and $f_\pi$ collectively as the infrared mass scale $\mlo$, OPE scales as
\begin{equation}
  V_{1\pi}(p', p) \sim \frac{1}{f_\pi^2} \sim \frac{4\pi}{m_N} \frac{1}{\mlo} \,,
\end{equation}
where $m_N \sim 4\pi f_\pi$ is used. The contact term $V_S^{(0)} = C_0$ enters also at LO. As argued in Ref.~\cite{Long:2012ve}, RG invariance requires a non-vanishing NLO generated by a momentum-dependent contact term, one order earlier than estimated from NDA,
\begin{equation}
  V_S^{(1)} = \frac{C_2}{2} \left({p'}^2 + p^2 \right) \sim \frac{4\pi}{m_N} \frac{1}{\mhi} \, .
\end{equation}

Even with this promotion, chiral EFT still converges slowly in $\cs{1}{0}$ (see, e.g., Ref.~\cite{Long:2012ve}). It was later proposed in Ref.~\cite{Long:2013cya} that the mass scale embedded in $C_2$ is actually quite small and can be identified with the inverse of the so-called generalized effective range $\simeq 130$ MeV that was first extracted in Ref.~\cite{Steele:1998zc}. This light mass scale needs to enter the LO amplitude, and promoting the $C_2$ term to LO is a choice to achieve this. The simplistic treatment of taking the sum of the $C_0$ and $C_2$ terms as the LO short-range potential has several obstacles, including satisfying RG invariance~\cite{Kaplan:1996xu,Beane:1997pk}
and overcoming the Wigner bound~\cite{Phillips:1996ae, Beane:1997pk}. However, these constraints do not necessarily mean that building a highly unnatural LO interaction with two adjustable parameter is a no-go scenario. Indeed, the dibaryon formalism used in Ref.~\cite{Long:2013cya} has been shown to be a viable implementation, making use of the kinetic energy of the dibaryon field to provide the much needed energy dependence of the potential.

On the other hand, no principle states that implementation of a two-parameter LO potential is unique. As long as a formulation incorporating both infrared scales satisfies the criteria introduced in Sec.~\ref{sec:intro}, it is acceptable. In fact, there is a clear practical motivation to search for other ways to improve the $\cs{1}{0}$ convergence: one encounters difficulties when applying the energy-dependent dibaryon potential~\eqref{eqn_Edep1s0} to many-nucleon calculations. It is the goal of the present paper to show that the separable, momentum-dependent alternative \eqref{eqn_pdep1s0} provides a satisfactory foundation to be expanded around upon.

We find it useful to cast the separable potential~\eqref{eqn_pdep1s0} into the form of a Lagrangian. An auxiliary dibaryon field $\bm{\phi}$ with spin-$0$ and isospin-$1$ is introduced, but $\bm{\phi}$ is stationary, namely, it does not have a kinetic term. The Lagrangian terms involving $\bm{\phi}$ are given by
\begin{equation}
\begin{split}
  \mathcal{L}_{\bm{\phi}} \, = \, \sigma\Delta \bm{\phi}^\dagger \cdot \bm{\phi} -\sqrt{\frac{4\pi}{m_N}} \sum_{m=1}^{3}\sum_{n = 0}^{\infty} \frac{g_{2n}}{2} \left[\phi_m^\dagger N^T \mathcal{P}_m\left(\frac{-\overleftrightarrow{\nabla}^2}{4m_N\Delta}\right)^n N + \mathrm{H.c.} \right]  + \cdots \,,
\end{split} \label{eqn:lagrangian}
\end{equation}
where
\begin{equation}
  \mathcal{P}_m=\frac{1}{\sqrt{8}} \tau_2\tau_m\sigma_2
\end{equation}
is the spin and isospin projector to ensure that $\bm{\phi}$ couples only to the $\cs{1}{0}$ channel and $\overleftrightarrow{\nabla}$ is defined so that
\begin{equation}
  N^T \overleftrightarrow{\nabla} N \equiv N^T (\overleftarrow{\nabla}-\overrightarrow{\nabla})N \, .
\end{equation}
Here $\sigma$ is normalized to $\pm 1$ by rescaling $\bm{\phi}$, and it will be determined by fitting to scattering data. If $g_{2n}$ are correlated by the binomial coefficients as
\begin{equation}
  \frac{g_{2n}}{g} = \binom{-\frac{1}{2}}{n} \, , \label{eq:binomal_gn}
\end{equation}
then we can obtain the $\phi NN$ vertex function after summing over all powers of relative momenta, 
\begin{equation}
  \mathcal{A}_{\phi NN} = -\frac{\sqrt{4\pi} g}{\sqrt{p^2/\Delta + m_N}} \mathcal{P}_m \, .
\label{eqn:phiNN_vertex}
\end{equation}
The desired separable $\cs{1}{0}$ potential then follows from the $s$-channel exchange of $\bm{\phi}$,
\begin{equation}
V_\text{spr}^{(0)}(p^\prime, p) = -\frac{4\pi}{m_N} \frac{\lambda}{\sqrt{p^{\prime 2} + m_N \Delta}\sqrt{p^2 + m_N \Delta}} \, ,
    \label{eqn:momdibpot}
\end{equation}
where $\lambda \equiv \sigma m_N g^2/4$.

In order to justify summing the $g_{2n}$ terms, $\lambda$ and $\Delta$ must be identified as two independent infrared mass scales,
\begin{equation}
    \sqrt{m_N \Delta} \sim \lambda \sim \mlo \, , \label{eqn:Delta_lambda_scale}
\end{equation}
which in turn translates into
\begin{equation}
    g \sim \, \sqrt{\frac{\mlo}{\mhi}}\, . \label{eqn:g_scale}
\end{equation}
We will see that the expected scaling of $\lambda$ and $\Delta$ is verified once their values are obtained from fitting to the $\cs{1}{0}$ empirical phase shifts.

Summing an infinite sequence of derivative coupling terms prompts concerns regarding chiral symmetry. Derivatives acting on the nucleon field must be accompanied by so-called chiral-connection operators involving the pion fields so that chiral symmetry is nonlinearly realized by these hadronic degrees of freedom~\cite{Coleman:1969sm,Callan:1969sn}. We will come back to this in Sec.~\ref{sec:chiral} to discuss chiral-connection operators of the $\phi NN$ transition vertex.

The LO $\cs{1}{0}$ amplitude is the resummation of $V^{(0)}_\text{spr}$ and $V_Y$ to all orders by way of the partial-wave LS equation,
\begin{equation}
   T(p^\prime, p; E) =  V_{\text{LO}}(p^\prime, p) + \frac{1}{2\pi^2} \int \mathrm{d} l \, l^2\, V_{\text{LO}}(p^\prime, l) \frac{T(l, p; E)}{E - l^2/m_N + i\epsilon}
   \label{eqn:LSE}
\end{equation}
with
\begin{align}
     V_{\text{LO}}(p^\prime, p) &= V_\text{spr}^{(0)}(p^\prime, p) + V_{Y}(p^\prime, p) \, , \\
     V_{Y}(p^\prime, p) &= -\frac{g_A^2}{4f_\pi^2} \frac{m_\pi^2}{q^2 + m_\pi^2} \, .
\end{align}
The LS equation is often schematically written as
\begin{equation}
  T^{(0)} = V_\text{LO} + V_\text{LO} G_0 T^{(0)} \, ,
\end{equation}
where $G_0$ is the nonrelativistic free-particle propagator. When solving the LS equation, we regularize the ultraviolet part of the potentials with a separable regulator to ensure that the regularization in one partial wave does not interfere another:
\begin{equation}
  V^\Lambda (\vec{p}\,^\prime,\vec{p})
  = f_R\!\left(\frac{p^{\prime\, 2}}{\Lambda^2}\right)
  \, V(\vec{p}\,^\prime,\vec{p}) \, f_R\!\left(\frac{p^2}{\Lambda^2}\right)
  \, . \label{eqn:Vgausreg}
\end{equation}
In particular, a Gaussian regulator is used in our numerical calculations:
\begin{equation}
  f_R(x) = e^{-x^2} \, .
\end{equation}

Using the two-potential method~\cite{Kaplan:1996xu}, the LO $\cs{1}{0}$ amplitude can be rewritten to facilitate analysis. We start by defining the off-shell Yukawa amplitude
\begin{equation}
    T_{Y}= V_Y + V_Y G_0 T_Y \, ,
\end{equation}
and the origin value of its scattering wave function, dressed by the square-root dipole form factor of the $\phi NN$ transition vertex~\eqref{eqn:phiNN_vertex}:
\begin{equation}
    \chi(p; E) = \frac{1}{\sqrt{p^2 + m_N\Delta}} + \int \frac{d^3 l}{(2\pi)^3} \left(l^2 + m_N \Delta\right)^{-\frac{1}{2}} \frac{T_Y(l, p)}{E - l^2/m_N + i\epsilon} \, .
\end{equation}
The LO amplitude is then given by
\begin{equation}
    T^{(0)}(k, k; E) = T_Y(k, k) - \frac{4\pi}{m_N} \frac{\chi(k; E)^2}{ \lambda^{-1} (k^2 + m_N\Delta) + I(E)} \, ,
\end{equation}
where
\begin{equation}
    I(E) \equiv \frac{4\pi}{m_N} \int  \frac{d^3 l}{(2\pi)^3} \left(l^2 + m_N \Delta\right)^{-\frac{1}{2}} \frac{\chi(l; E)}{E - l^2/m_N + i\epsilon} \, .
\end{equation}
Compared with the energy-dependent dibaryon potential, the factor $(p^2 + m_N \Delta)^{-\frac{1}{2}}$ brings so much UV suppression that the integral would be finite even without the regulator, and $\Delta(\Lambda)$ and $\lambda(\Lambda)$ approach finite values for $\Lambda \to \infty$. Table~\ref{tab_ydelta} shows their values as a function of $\Lambda$. The fitting to empirical $\cs{1}{0}$ phase shifts will be explained in Sec.~\ref{sec:results}. Using these values, we have $\sqrt{m_N \Delta} \simeq 150$ MeV and $\lambda \simeq 200$ MeV, which confirms the expected scaling given in Eq.~\eqref{eqn:Delta_lambda_scale}. The inverse of $\sqrt{m_N \Delta}$ comes out close to the pion Compton wave length $\simeq 1.4$ fm, suggesting that the dibaryon potential has a range similar to that of OPE.

\begin{table}[tb]
\caption{
Running of $\lambda $ (MeV) and $\Delta$ (MeV) with $\Lambda$ (MeV) at LO.
}
\centering
\begin{tabular}{ccc}
\hline
\hline
$\Lambda$\; & $\Delta^{(0)}$ \; & $\lambda^{(0)}$\\
\hline
600 & 27.1 & 270 \\
% 800 & 25.5 & 243 \\
1200 & 24.6 & 223 \\
2400 & 24.1 & 209 \\
\hline
\hline
\end{tabular}
\label{tab_ydelta}
\end{table}

Let us turn to higher orders. Although $\lambda(\Lambda)$ and $\Delta(\Lambda)$ appear at LO, their running with $\Lambda$ can be modified at subleading orders:
\begin{align}
 \lambda(\Lambda) &= \lambda^{(0)}(\Lambda) + \lambda^{(1)}(\Lambda) + \lambda^{(2)}(\Lambda) + \cdots \, , \\
 \Delta(\Lambda) &= \Delta^{(0)}(\Lambda) + \Delta^{(1)}(\Lambda) + \Delta^{(2)}(\Lambda) + \cdots \, .
\end{align}
The $\lambda^{(\nu)}$ and $\Delta^{(\nu)}$ are formally smaller than their LO value by $Q^\nu/\mhi^\nu$. These modifications play a role at higher orders, which can be constructed by the generating function
\begin{equation}
    \mathcal{F}_\text{spr}(p', p; x) \equiv -\frac{4\pi}{m_N} \frac{\lambda(x)}{\sqrt{p'^2 + m_N \Delta(x)}\sqrt{p^2 + m_N \Delta(x)}}
    \, ,
\end{equation}
where $x$ is an auxiliary variable to generate the expansion and
\begin{align}
  \lambda(x) &= \lambda^{(0)} + x \lambda^{(1)} + x^2 \lambda^{(2)} + \cdots \, , \\  \Delta(x) &= \Delta^{(0)} + x \Delta^{(1)} + x^2 \Delta^{(2)} + \cdots   \, .
\end{align}
From there, the dibaryon part at each order is generated by
\begin{equation}
    \mathcal{F}_\text{spr}(p', p; x) = V^{(0)}_\text{spr}(p', p) + x V^{(1)}_\text{spr}(p', p) + x^2 V^{(2)}_\text{spr}(p', p) + \cdots \, ,
    \label{eqn:GeneratingFunc}
\end{equation}
with the first two corrections explicitly given by
\begin{equation}
  \begin{split}
  V^{(1)}_\text{spr}(p', p) \, =\, & -\frac{4\pi}{m_N} \frac{1}{\sqrt{p^2 + m_N\Delta^{(0)}}\sqrt{p^{\prime 2} + m_N\Delta^{(0)}}} \\
  &\times  \left\{\lambda^{(1)} - \frac{m_N \lambda^{(0)}}{2} \left(\frac{1}{p^2 + m_N\Delta^{(0)}} + \frac{1}{p^{\prime 2} + m_N\Delta^{(0)}} \right)\Delta^{(1)} \right\}
  \end{split}
\end{equation}
and
\begin{equation}
  \begin{split}
    V^{(2)}_\text{spr}(p', p) \, &=\, -\frac{4\pi}{m_N} \frac{1}{\sqrt{(p^2 + m_N\Delta^{(0)})}\sqrt{(p^{\prime2} + m_N\Delta^{(0)})}} \\
    & \times \left\{ \lambda^{(2)}  - \frac{m_N}{2}\left[ \frac{1}{(p^2 + m_N\Delta^{(0)})} + \frac{1}{(p^{\prime2} + m_N\Delta^{(0)})} \right] \left( \lambda^{(0)}\Delta^{(2)} + \lambda^{(1)}\Delta^{(1)} \right) \right. \\
    & \qquad + \frac{m_N^2 \lambda^{(0)}}{8} \left[ \frac{3}{(p^2 + m_N\Delta^{(0)})^{2}} + \frac{2}{(p^2+ m_N\Delta^{(0)})(p^{\prime 2} + m_N\Delta^{(0)})}  \right.\\
    & \quad \qquad \qquad \qquad \left. \left. + \frac{3}{(p^{\prime2} + m_N\Delta^{(0)})^{2}} \right] \Delta^{(1)^2} \right\}  \, .
  \end{split}
\end{equation}

If one expands the potential~\eqref{eqn:momdibpot} in powers of momenta, a series of $\cs{1}{0}$ contact terms emerges. Therefore, the values of $\Delta$ and $\lambda$ span a surface in the parameter space of $\cs{1}{0}$ contact couplings. The deviation away from the surface can be described by residual values of $\cs{1}{0}$ contact terms. Reference~\cite{Long:2013cya} has shown that higher-order short-range forces are parametrized by conventional four-nucleon operators
\begin{equation}
    V_S = \sum_{n = 0}^\infty \frac{C_{2n}}{2} \left({p'}^{2n} + {p}^{2n}\right)
    \label{eqn:C2n_def}
\end{equation}
with the following scaling for $C_{2n}$:
\begin{equation}
    C_{2n} \sim \frac{4\pi}{m_N} \frac{1}{\mlo^{n} \mhi^{n+1}} \, .
\end{equation}

NLO and higher-orders amplitudes are built as perturbations on top of LO, i.e., they are diagrammatically perturbative insertions of $V_\text{NLO}$,  $V_{\text{N}^2\text{LO}}$ (and so on) into the LO amplitude:
\begin{equation}
\begin{split}
  T^{(1)} &= (1 + T^{(0)}G_0) V_{\text{NLO}} (G_0T^{(0)} + 1) \, ,\\
  T^{(2)} &= (1 + T^{(0)}G_0) [V_{\text{N}^2\text{LO}} + V_{\text{NLO}} G_0(1 + T^{(0)}G_0) V_{\text{NLO}}] (G_0T^{(0)} + 1) \, , \\
  & \cdots  \, .
\end{split}
\end{equation}
In particular, since there is no pion-exchanges at NLO, the $\cs{1}{0}$ potential is given by
\begin{equation}
    V_\mathrm{NLO}(p', p) = C_0^{(0)} + V^{(1)}_\mathrm{spr}(p^\prime, p) \, .
\end{equation}

\section{Radiative corrections\label{sec:Radpi}}

\begin{figure}
  \centering
  \includegraphics[scale=0.9]{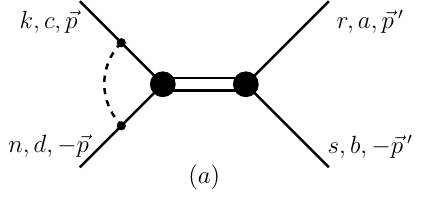}
  \hspace{10pt}
  \includegraphics[scale=0.9]{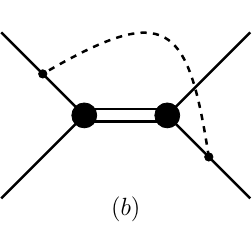}
  \caption{Radiation-pion Feynman diagrams contributing to {\NNLO} potentials. The solid (dashed) lines represent nucleons (pions), and the solid blobs are the
  $\phi NN$ vertex function~\eqref{eqn:phiNN_vertex}.
(a) The pion line connects both nucleons either in the initial or in the final states. (b) The pion line connects one of the incoming nucleon lines to one of the final nucleon legs. Other variants due to permuting the pion line on different nucleon external lines are not shown.
}
\label{fig:radpi_diagram}
\end{figure}

The scaling \eqref{eqn:Delta_lambda_scale} has so far led us to have the momentum-dependent dibaryon potential (and its subleading corrections) simply substitute the energy-dependent one. But there is an important difference at {\NNLO}. Besides the leading two-pion exchange $V_{\mathrm{TPE0}}$~\cite{Kaiser:1997mw}, pionic radiative corrections to the dibaryon potential, depicted in Fig.~\ref{fig:radpi_diagram}, appear at this order. For a dynamically propagating dibaryon, the radiative corrections are zero-range interactions (much like the radiative corrections to pure $NN$ contact operators~\cite{Ordonez:1995rz, Epelbaum:1999dj}); therefore, they do not change the form of the potentials in any nontrivial way. For the momentum-dependent dibaryon vertex, this is no longer the case because the vertex is finite-ranged.

In Fig.~\ref{fig:radpi_diagram}, $\vec{p}$ ($\vec{p}\,'$) is the incoming (outgoing) nucleon c.m. momentum, and $k$, $n$, $r$, and $s$ ($a$, $b$, $c$, and $d$) are spin (isospin) indexes. While Fig.~\ref{fig:radpi_diagram} (a) contributes only to $\cs{1}{0}$, the impact of Fig.~\ref{fig:radpi_diagram} (b) is confined to the triplet channels, as will be shown later in the section.
Without the dibaryon field, the convergence in the triplet channels is already quite good~\cite{Long:2011xw, PavonValderrama:2011fcz}. Thus, it is important for the purpose of validating the dibaryon potential to examine how it affects the triplet channels through radiative corrections. If Fig.~\ref{fig:radpi_diagram} (b) interferes with the triplet channels so much that a satisfactory description of data is spoiled, we will have to give up the momentum-dependent dibaryon formalism even if it does improve $\cs{1}{0}$. Fortunately, it turns out that these radiative corrections for the most part only renormalize the $NN$ contact terms in $\csd$ and do not significantly change the on-shell amplitudes. Before turning to the phase shifts, we discuss calculation of the Feynman diagrams in Fig.~\ref{fig:radpi_diagram}.

As mentioned previously, the diagram of Fig.~\ref{fig:radpi_diagram} (a) is expected to contribute only to $\cs{1}{0}$, for it can be broken by cutting the dibaryon propagator. The $NN$ reducible part of the diagram is the contribution from picking up the nucleon pole when integrating out the zeroth component of the loop momentum, which is part of the LO amplitude and has been accounted for when iterating $V_\text{LO}$ in Eq.~\eqref{eqn:LSE}. What is counted as {\NNLO} is the irreducible part of the diagram that results from picking up the pion pole and is given by the three-dimensional integral
\begin{equation}
\begin{split}
& \mathcal{A}_\text{rad}^{(a)} =
-\frac{4\pi}{m_N} \frac{g_A^2}{4 f_\pi^2} \lambda^{(0)}(\Lambda) \sum_{m}\left[(\mathcal{P}_m^\dagger)_{rs,ab}(\mathcal{P}_m)_{kn,cd} \right] \\
& \quad \times \int \frac{\mathrm{d}^3l}{(2\pi)^3} \; \frac{l^2}{\left(l^2 + m_\pi^2\right)^{\frac{3}{2}}}\, \mathcal{R}(\left|\vec{p}\,'\right|, |\vec{p}-\vec{l}\,|; \Lambda) \, ,
\end{split}
\end{equation}
with the scalar function
\begin{equation}
    \mathcal{R}(x, y; \Lambda) \equiv \frac{ f_R\left(\frac{x^ 2}{\Lambda^2}\right) f_R\left(\frac{y^2}{\Lambda^2}\right)}{ \sqrt{x^2 + m_N\Delta^{(0)}(\Lambda)} \sqrt{y^2 + m_N\Delta^{(0)}(\Lambda)}} \, .
\end{equation}
$\lambda^{(0)}(\Lambda)$ and $\Delta^{(0)}(\Lambda)$ are the LO values, determined by fitting to empirical $\cs{1}{0}$ phase shifts (see Table~\ref{tab_ydelta}). On the ground of satisfying unitarity up to this order, one must use the same regularization function adopted for the LO resummation~\eqref{eqn:Vgausreg}.

The amplitude of the diagram in which the pion propagator connects the outgoing nucleon lines (not shown in Fig.~\ref{fig:radpi_diagram}) can be obtained by making the following replacements:
\begin{equation}
    (a,b)(c,d) \rightarrow (c,d)(a,b) \; ; \quad (r,s)(k,n) \rightarrow (k,n)(r,s) \;; \quad (\vec{p}, \vec{p}\,^\prime) \rightarrow (-\vec{p}\,^\prime, -\vec{p}\,) \, .
\end{equation}
The integrals are calculated numerically, and the trace formalism of Ref.~\cite{Fleming:1999ee} is used to obtain partial-wave projected amplitudes before the loop momentum is integrated over. With a bit of spin-isospin algebra, the contribution of Fig.~\ref{fig:radpi_diagram} (a) and its permutation to $\cs{1}{0}$ is obtained by evaluating the following integral:
\begin{equation}
\begin{split}
    V_\text{rad}^{\cs{1}{0}} &= -\frac{\pi}{m_N} \frac{g_A^2}{4 f_\pi^2} \lambda^{(0)}(\Lambda) \int \frac{\mathrm{d}^3l}{(2\pi)^3} \frac{l^2}{\left(l^2 + m_\pi^2\right)^{\frac{3}{2}}} \\
    & \quad \times \left[ \mathcal{R}(\left|\vec{p}\,'\right|, |\vec{p}-\vec{l}\,|; \Lambda) + \mathcal{R}(\left|\vec{p}\,\right|,|\vec{p}\,^\prime + \vec{l}\,|; \Lambda) \right] \, .
\end{split}
    \label{eqn:Vrad1s0}
\end{equation}

The corrections to the triplet channels come from the diagrams in Fig.~\ref{fig:radpi_diagram} (b),
\begin{equation}
\begin{split}
   \mathcal{A}_\text{rad}^{(b)} &= \frac{4\pi}{m_N} \frac{g_A^2}{16f_\pi^2}  \lambda^{(0)}(\Lambda) \sum_{e, m, \alpha, \beta } \frac{1}{8} \left[ (\tau_e \tau_2 \tau_m)_{ba}(\tau_2 \tau_e \tau_m)_{cd} (\sigma_\alpha \sigma_2)_{rs} (\sigma_2 \sigma_\beta)_{kn} \right] \\
   &\quad \times \int \frac{\mathrm{d}^3l}{(2\pi)^3} \frac{l_\alpha l_\beta}{\left(l^2 + m_\pi^2 \right)^{\frac{3}{2}}}\, \mathcal{R}\left(\left|\vec{p}_-\right|, \left|\vec{p}\,^\prime_+ \right|; \Lambda\right) \,,
\end{split}
\end{equation}
where
\begin{equation}
  \vec{p}_\pm \equiv \vec{p} \pm \frac{\vec{l}}{2}, \qquad \vec{p}\,^\prime_\pm \equiv \vec{p}\,^\prime \pm \frac{\vec{l}}{2} \, .
\end{equation}
Projection of $\mathcal{A}_\text{rad}^{(b)}$ is again performed with the trace formalism. After taking all the variants due to permutations of the pion line into account, the radiative corrections to the $\csd$ potential is given by
\begin{equation}
\begin{split}
     V_\text{rad}^{\csd}  = & -\frac{\pi}{m_N} \frac{g_A^2}{16f_\pi^2}  \lambda^{(0)}(\Lambda) \int \frac{\mathrm{d}^3l}{(2\pi)^3} \frac{l^2}{\left(l^2 + m_\pi^2\right)^{\frac{3}{2}}} \mathcal{T}(\hat{p},\hat{p}^\prime, \hat{l}) \\
     \times & \left[ \mathcal{R} \left( \left|\vec{p}_-\right|, \left|\vec{p}\,^\prime_+ \right|; \Lambda \right) + \mathcal{R} \left( \left|\vec{p}_{-}\right|, \left|\vec{p}\,^\prime_- \right|; \Lambda \right) \right.
    \\
     & + \left.\mathcal{R} \left( \left|\vec{p}_+ \right|, \left|\vec{p}\,^\prime_- \right|; \Lambda \right) + \mathcal{R} \left( \left|\vec{p}_+\right|, \left|\vec{p}\,^\prime_-\right|; \Lambda \right) \right]
     \,, \label{eqn:Vrad3s1}
\end{split}
\end{equation}
where the $2\times 2$ matrix $\mathcal{T}(\hat{p},\hat{p}^\prime, \hat{l})$ is defined by
\begin{equation}
\begin{split}
    \langle \cs{3}{1} | \mathcal{T} | \cs{3}{1} \rangle &= 1 \, ,\\
    \langle \cs{3}{1} | \mathcal{T} | \cd{3}{1} \rangle &= \frac{3(\hat{p} \cdot \hat{l})^2}{\sqrt{2}} - \frac{1}{\sqrt{2}} \, , \\
    \langle \cd{3}{1} | \mathcal{T} | \cs{3}{1} \rangle &= \frac{3(\hat{p}^\prime \cdot \hat{l})^2}{\sqrt{2}} - \frac{1}{\sqrt{2}} \, , \\
   \langle \cd{3}{1} | \mathcal{T} | \cd{3}{1} \rangle &= \frac{9}{2}(\hat{p} \cdot \hat{l})(\hat{p}^\prime \cdot \hat{l})\, \hat{p}\cdot\hat{p}^\prime -\frac{3}{2}\left[(\hat{p}^\prime \cdot \hat{l})^2 + (\hat{p} \cdot \hat{l})^2\right] + \frac{1}{2} \, .
\end{split}
\end{equation}

\section{Nucleon-nucleon phase shifts and triton binding energy\label{sec:results}}

In this section we show how well the momentum-dependent dibaryon potential agrees with empirical phase shifts of $NN$ scattering and its prediction for the triton binding energy at LO and NLO.

\subsection{\texorpdfstring{${}^{\boldsymbol{1}}\boldsymbol{S}_{\boldsymbol{0}}$}{1S0} channel}

\begin{figure}[tb]
  \centering
  \includegraphics[scale=0.5]{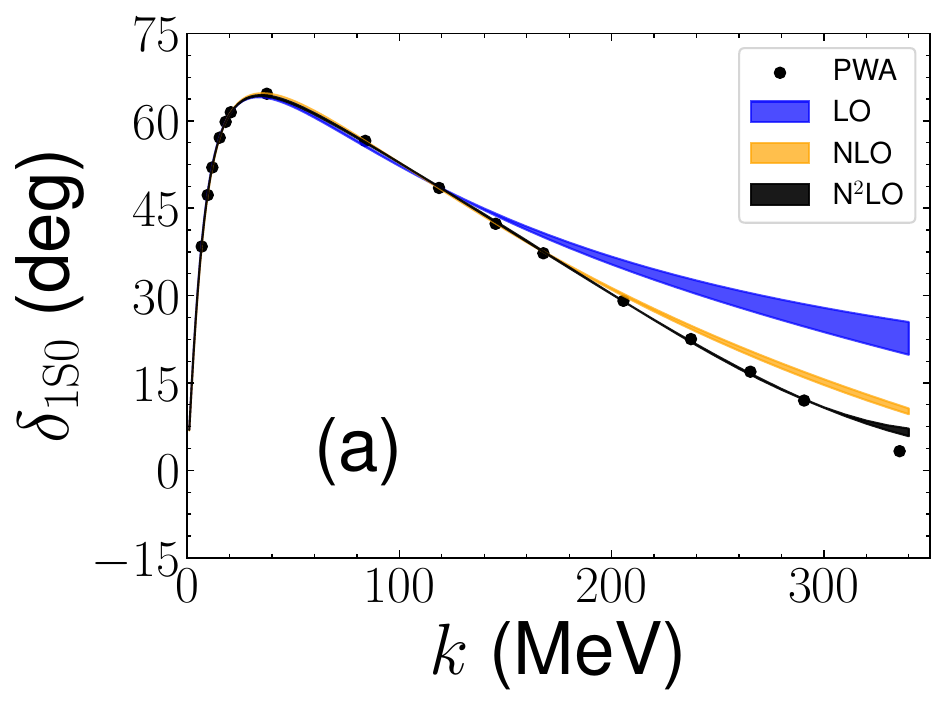}
  \includegraphics[scale=0.5]{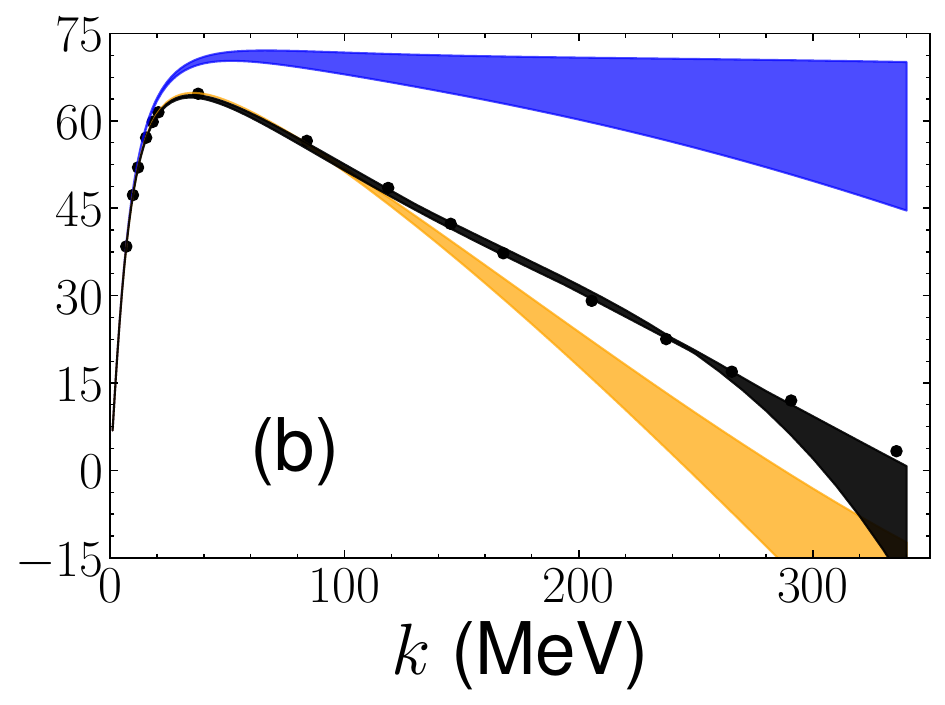}
  \includegraphics[scale=0.5]{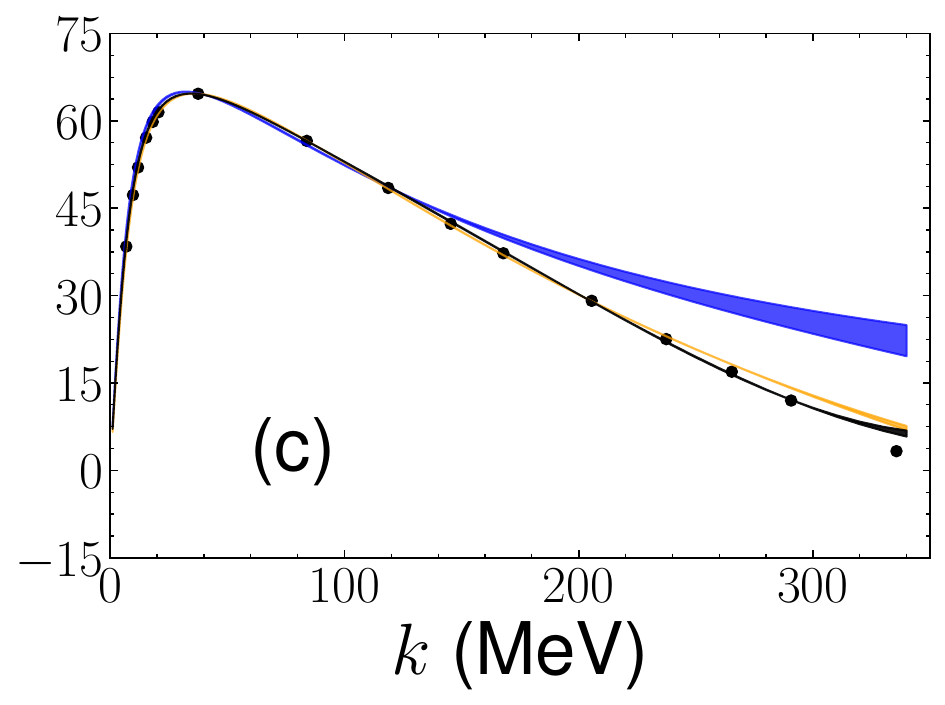}
  \caption{The $^1S_0$ phase shifts as a function of c.m. momentum $k$ up to {\NNLO}.  The blue, orange and black bands represent LO, NLO, and {\NNLO} respectively, and the bands show the variation with $\Lambda$ from $600$ to $2400$ MeV.
  The potentials used in (a), (b) and (c) are momentum-dependent dibaryon potentials, MMW scheme and dynamic dibaryon potentials respectively.
}
  \label{fig:1s0_phase}
\end{figure}

Having computed the radiative corrections, we write the {NNLO} $\cs{1}{0}$ potential explicitly as follows:
\begin{equation}
    V_{\text{N}^2\text{LO}} = C_0^{(2)} + \frac{C_2^{(0)}}{2}(p^2 + p^{\prime 2}) + V^{(2)}_{\mathrm{spr}}(p^\prime, p) + V_{\mathrm{TPE0}} + V_\text{rad}^{\cs{1}{0}}(p^\prime, p) \, .
\end{equation}
The $\cs{1}{0}$ phase shifts up to N$^2$LO are shown in Fig.~\ref{fig:1s0_phase} (a).  The bands represent the cutoff variation spanning from $\Lambda = 600$ MeV to 2400 MeV. The LECs are determined by fitting the EFT amplitudes to the empirical phase shifts provided by the partial-wave analysis (PWA) of the Nijmegen group~\cite{NNonline, Stoks:1993tb} by a least-square procedure. For LO and NLO, PWA points with c.m. momentum $k \leqslant 150$ MeV are used, and at {\NNLO} inputs from $k = 150 \sim 300$ MeV are added. For comparison, the phase shifts up to {\NNLO} obtained with the dynamic dibaryon potentials of Ref.~\cite{Long:2013cya} and the power counting of Ref.~\cite{Long:2012ve}---referred to as minimally modified Weinberg (MMW) in the current paper---are shown in Fig.~\ref{fig:1s0_phase} (b) and (c). The LECs for this interaction were determined with the same set of PWA inputs.

The rather mild sensitivity to the cutoff $\Lambda$, manifest in the narrowness of the cutoff-variation bands, is somewhat expected based on the ultraviolet suppression afforded by the momentum-dependent $\phi NN$ vertex. The rapid order-by-order convergence is also retained from the energy-dependent dibaryon potential, as illustrated in Ref.~\cite{Long:2013cya}. Moreover, the radiative correction $V_\text{rad}^{\cs{1}{0}}$ does not spoil the cutoff independence or the convergence of the EFT expansion.

\subsection{Spin-triplet channels}

$\csd$ is the spin-triplet channel that is most likely affected by the pion cloud because the centrifugal barrier in higher waves tends to suppress the radiated pions. In calculating the phase shifts of $\csd$, we follow the power counting of Ref.~\cite{Long:2011xw}, with OPE and one $S$-wave counterterm at LO, NLO vanishing, and two more counterterms, TPE0, and $V_\text{rad}^{\csd}$~\eqref{eqn:Vrad3s1} at {\NNLO}.

\begin{figure}
  \centering
  \includegraphics[scale=0.45]{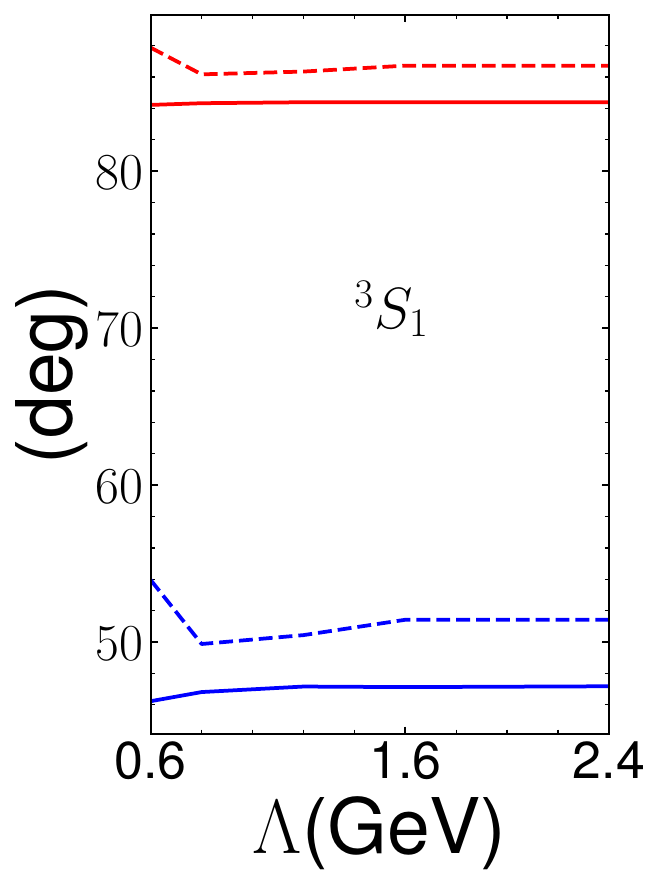}
  \includegraphics[scale=0.45]{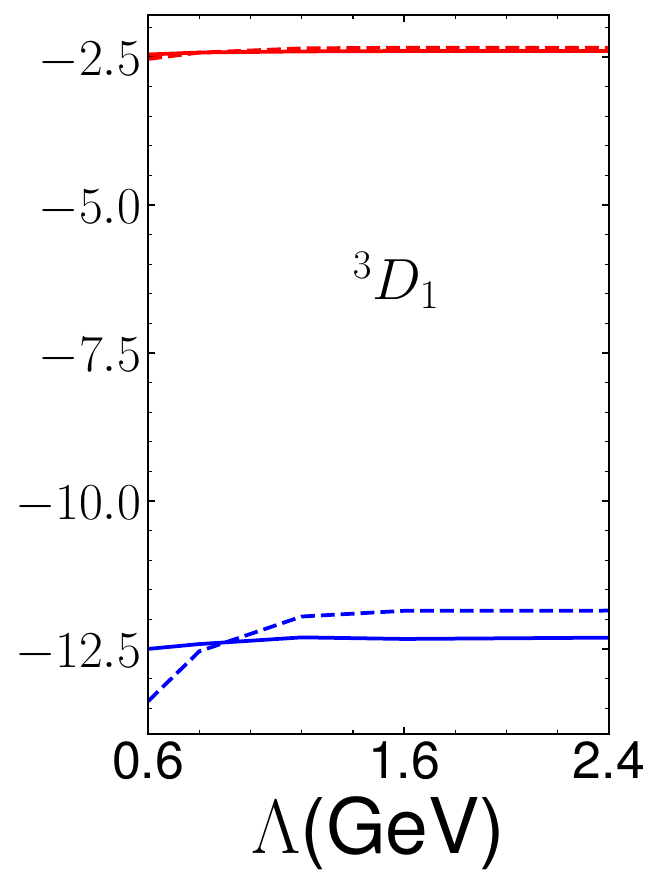}
  \includegraphics[scale=0.45]{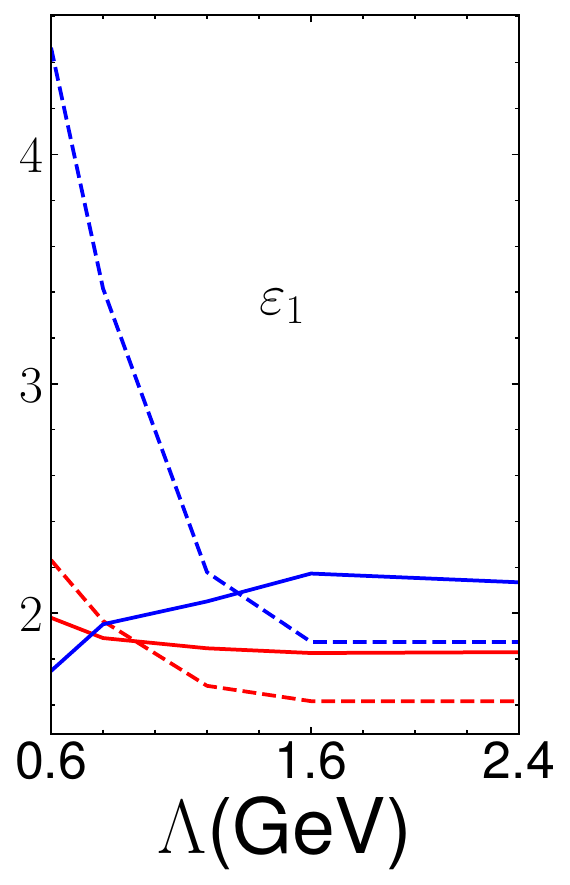}
  \caption{The $\csd$ phase shifts and mixing angle at $k = 100$ MeV (red) and $200$ MeV (blue), as functions of momentum cutoff $\Lambda$. The dashed (solid) lines represent the LO ({\NNLO}) result.
  \label{fig_N2_201_cut}}
\end{figure}

The cutoff dependence has typically been more of a concern for the triplet channels because the singular attraction of OPE was a major source to upset NDA from the perspective of RG invariance. We determine the values of LECs by demanding that several PWA phase shifts are reproduced exactly. Specifically, the following PWA phase shifts are used: the $\cs{3}{1}$ phase shift $\delta_{^3S_1}$ at $k = 118$ MeV and $153$ MeV and the mixing angle $\epsilon_1$ at $153$ MeV. Shown in Fig.~\ref{fig_N2_201_cut}, $\csd$ phase shifts and mixing angles for representative momenta ($k = 100$ and 200 MeV) become insensitive to the cutoff value $\Lambda$ when $\Lambda$ is sufficiently large.

\begin{figure}
  \centering
  \includegraphics[scale=0.45]{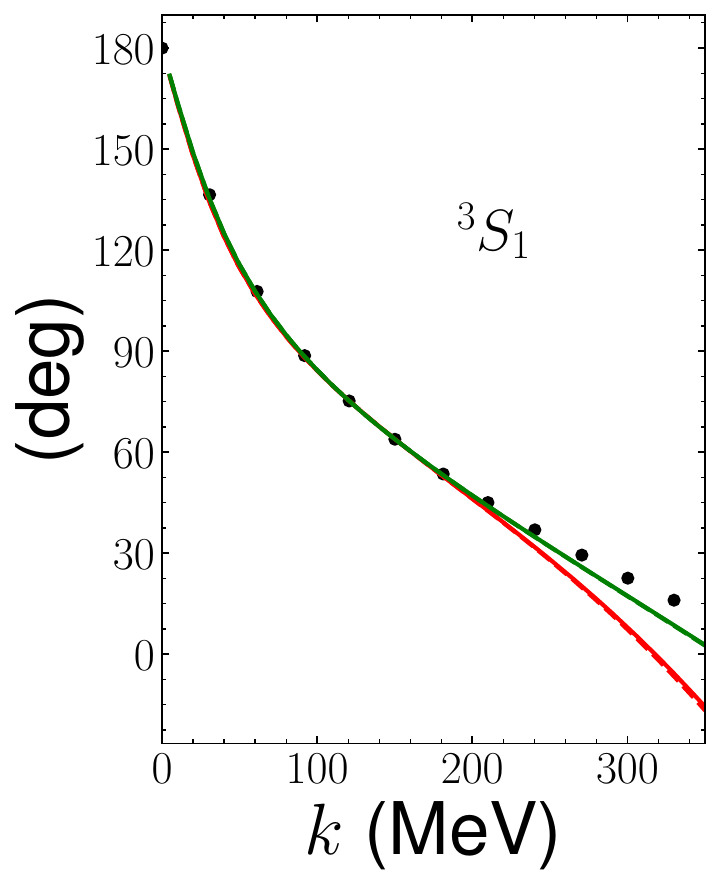}
  \includegraphics[scale=0.45]{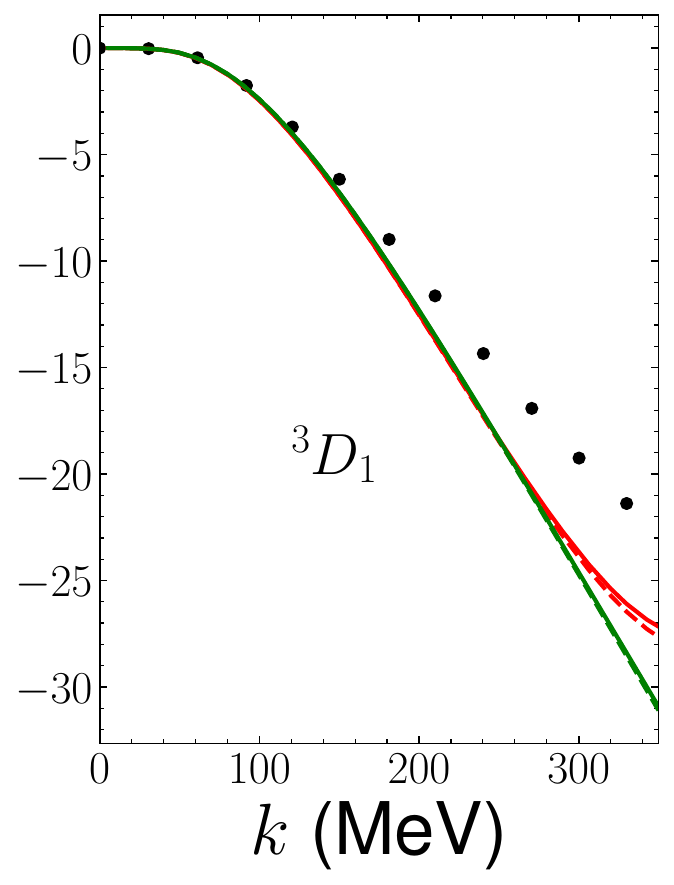}
  \includegraphics[scale=0.45]{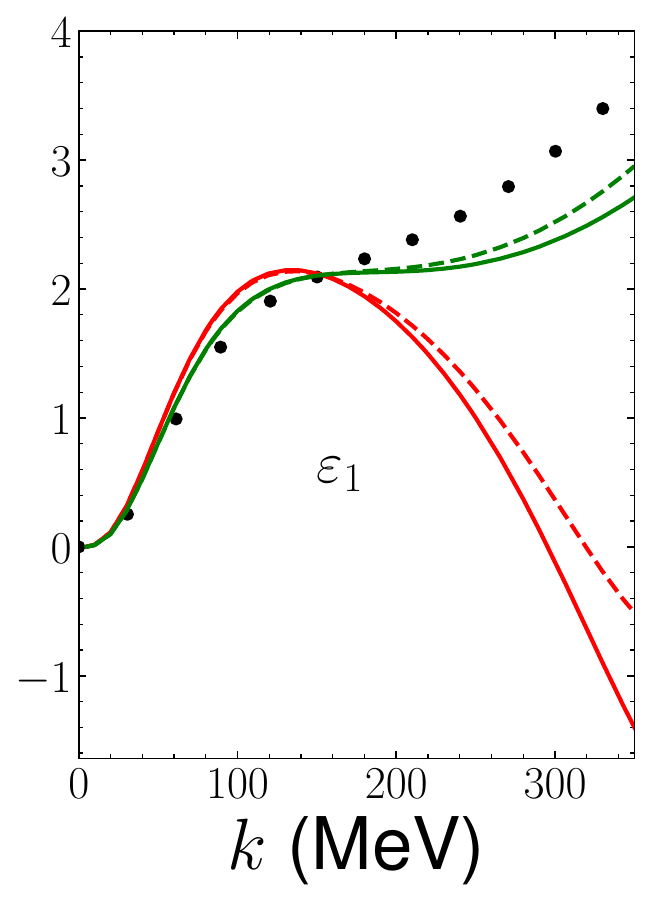}
  \caption{The $\csd$ phase shifts and mixing angle as functions of the c.m. momentum $k$, for cutoff $\Lambda = 600$ MeV (red) and $2400$ MeV (green). The solid lines represent the results with radiative corrections to the dibaryon potential, and the dashed lines correspond to the MMW scheme. The circles are the PWA values. See the text for more detailed explanation.
}
  \label{fig_N2_201_phase_NRM}
\end{figure}

Figure~\ref{fig_N2_201_phase_NRM} compares the $\csd$ phase shifts produced by the radiatively-corrected dibaryon potential and the MMW scheme. The difference is minuscule: for the $\cs{3}{1}$ phase shifts, the discrepancy is no more than about one degree, whereas the differences of the $\cd{3}{1}$ phase shifts and the mixing angle is smaller than one degree.

However, the values of $\csd$ counterterms are renormalized by these radiative corrections, varying up to $50\%$ compared to their MMW values. Thus, the impact of the radiative corrections on the triplet channels appears to be almost entirely short-ranged. This is further confirmed by the numerically tiny contributions to higher partial waves such as $^3P_0$, $^3P_1$, and $^3D_2$. We have computed the phase shifts of these partial waves, using partially-perturbative-pion power counting, Ref.~\cite{Long:2011qx} for $\cp{3}{0}$ and Ref.~\cite{Wu:2018lai} for $\cp{3}{1}$ and $\cd{3}{2}$. Within numerical precision of our calculations, the differences are found to be negligible.

\subsection{Triton binding energy\label{sec:triton}}

The success of the momentum-dependent dibaryon potential in $\cs{1}{0}$ was more or less expected because it is unitarily equivalent to the energy-dependent version in $\cs{1}{0}$, at least at tree level. A third particle can assess its impact beyond the two-nucleon system. The exchanged pion in the radiative corrections considered in Sec.~\ref{sec:Radpi} plays the role of a third particle, but it is virtual. We therefore now asses what the theory predicts for the triton binding energy up to NLO.

The calculation is set up following the power counting of partially perturbative pions. At LO, we iterate to all orders the potential~\eqref{eqn:momdibpot} plus OPE in $\cs{1}{0}$, and one contact term plus OPE in $\csd$ and $\cp{3}{0}$~\cite{Nogga:2005hy}. This is done by solving the Faddeev equations to obtain binding energies as LO. At NLO, in addition to the NLO $\cs{1}{0}$ force, OPE is included in channels up to $D$ waves ($\cp{1}{1}$, $\cp{3}{1}$, $\cpf$, $\cd{1}{2}$, $\cd{3}{2}$, and $\cdg$). Perturbative corrections to the binding energy are obtained by evaluating the NLO potentials between LO wavefunctions, obtained from the Faddeev equations along with the LO energies. For all calculations we include channels with angular momentum up to $j_{\text{max}}=4$ in the Faddeev equations.

The results for various cutoff are shown in Table~\ref{tab:bd_triton} as columns of labeled ``SEP.'' For comparison, we also show the results for the MMW scheme. Since the triton is relatively shallow, with an average binding momentum $\sqrt{2m_N B_\text{3H} / 3} \simeq 73$ MeV, we do not expect the dibaryon potential to necessarily perform much better in reproducing the triton binding energy. In fact, the NLO of MMW has a smaller cutoff variation, and the large $\Lambda$ limit is closer to the experimental value of $\simeq 8.5$ MeV. However, cutoff variation at best should be taken as a lower bound for the full theoretical uncertainty. The NLO correction of SEP is much smaller than MMW, which might imply that the actual EFT truncation error is smaller with SEP than with MMW. While a more systematic comparison of the two schemes in few- or many-nucleon is definitely interesting, it is beyond the scope of the paper, as is a triton calculation involving the energy-dependent dibaryon potential. We are for now content with finding that the the dibaryon potential gives a description of the three-nucleon system that is comparable to the MMW scheme. Both the SEP and MMW interactions yield the triton slightly more bound compared to earlier calculations using a similar perturbative power-counting scheme~\cite{Song:2016ale,Song:2016ale-err}.

Nuclear-structure calculations using the dibaryon interaction~\cite{Yang:2020pgi} or an interaction inspired by it~\cite{SanchezSanchez:2020kbx} have been carried out previously. In Ref.~\cite{SanchezSanchez:2020kbx}, some of the light nuclei with $A = 3 \sim 6$ were investigated. A direct comparison between the triton binding energy calculated in Ref.~\cite{SanchezSanchez:2020kbx} and this work is however difficult because different power-counting and regularization schemes have been used. We only note that results agree within $10\%$.

\begin{table}
\caption{The binding energy of the triton (MeV). See the text for more explanations.}
\centering
\begin{tabular}{ccccccccc}
\hline
\hline
 & & \multicolumn{3}{c}{SEP} & & \multicolumn{3}{c}{MMW} \\
 \cline{3-5} \cline{7-9}
 $\Lambda$ (MeV) & & LO & & NLO & & LO & & NLO \\
\hline
  400 & &  -8.67 & &  -8.50 & &  -11.02 & &  -7.68 \\
  600 & &  -6.10 & &  -6.10 & &   -6.46 & &  -6.90 \\
  800 & &  -5.57 & &  -5.58 & &   -5.30 & &  -6.64 \\
  1600 & &  -5.37 & &  -5.51 & &   -4.89 & &  -6.49 \\
\hline
\hline
\end{tabular}
\label{tab:bd_triton}
\end{table}

\section{Chiral symmetry\label{sec:chiral}}

Nonlinear realization of spontaneously broken chiral symmetry requires derivatives acting on hadronic fields to be chirally covariant. The chiral-covariant derivative $\mathscr{D}_i$ for the nucleon, which is an isospin-$1/2$ field, is given as~\cite{Weinberg:1996kr}
\begin{equation}
  \mathscr{D}_i N \equiv \left(\nabla_i + \frac{\bm{\tau}}{2}\cdot \bm{E}_i \right)N \,,
  \label{eq:covdervt}
\end{equation}
where the so-called chiral connection $\bm{E}_i$ is given by
\begin{equation}
   \bm{E}_i \; \equiv \; i\frac{\bm{\pi}}{f_\pi}\times \bm{D}_i\,,
\end{equation}
with
\begin{equation}
   \bm{D}_i \equiv D^{-1} \frac{\nabla_i \bm{\pi}}{2f_\pi} \qquad \text{and} \qquad D \equiv 1+\frac{\bm{\pi}^2}{4 f_{\pi}^{2}} \,.
\end{equation}
The moral here is that derivatives acting on $N$ must be accompanied by composite chiral-connection terms consisting of $\bm{\pi}$ and $N$, with two types of operators sharing the same set of LECs. Because the $\phi NN$ vertex function~\eqref{eqn:phiNN_vertex} is constructed out of an infinite series of derivative couplings acting on $N$, it is less straightforward than usual to write down the corresponding chiral-connection operators. Furthermore, in light of the proposed two-parameter fine-tuning at LO in $\cs{1}{0}$, these chiral-connection operators could be enhanced in comparison with NDA.

\begin{figure}
  \centering
  \includegraphics[scale=0.8]{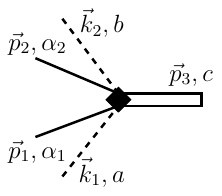}
  \caption{The transition vertex of $NN \pi \pi \to \bm{\phi}$.
  \label{fig:phiNNpipi}}
\end{figure}

Chiral-connection operators that include products of two pion fields, the minimal number of pion fields such operators expected to have, are normally the most phenomenologically important for low-energy nuclear physics. We therefore focus on the transition vertex of $NN \pi \pi \to \bm{\phi}$, depicted in Fig.~\ref{fig:phiNNpipi}, where $\alpha_{1, 2}$ denotes collectively the spin and isospin indices of the incoming nucleons, $\vec{p} \equiv (\vec{p}_1 - \vec{p}_2)/2$ their relative momentum, $a,b,c$ the isospin indices of the pions and the isovector dibaryon, and $\vec{k}_{1, 2}$ the momentum of the incoming pions. Following the derivations in the Appendix, one finds the vertex function of $\phi NN\pi \pi$ to be
\begin{equation}
  \begin{split}
  \mathcal{A}_{\phi NN\pi\pi} = \frac{ig}{4 f_{\pi}^{2}} \sqrt{\frac{4\pi}{m_N}} &\biggl\{ i \left(\delta_{bc}\mathcal{P}_a - \delta_{ac}\mathcal{P}_b \right)_{\alpha_2\alpha_1} \mathcal{B}_{+}
 \\
    & \quad \null+
     \frac{1}{\sqrt{8}} \epsilon_{abc} (\sigma_{2} \tau_{2})_{\alpha_2\alpha_1} \mathcal{B}_{-}
 \biggr\} \,,
  \label{eqn:AphiNNpipi}
  \end{split}
\end{equation}
where we have defined
\begin{equation}
\begin{split}
\mathcal{B}_\pm &\equiv u\left(\left| \vec{p} + \vec{k}_1/2 \right|, \left|\vec{p} + \vec{k}_2/2 \right| \right) \\
&\quad \pm u\left(\left| \vec{p} - \vec{k}_1/2 \right|, \left|\vec{p} - \vec{k}_2/2 \right|\right)
\, , \\
  u(x, y) &\equiv \left(1 + \frac{x^2}{m_N\Delta} \right)^{-\frac{1}{2}} - \left(1 + \frac{y^2}{m_N\Delta} \right)^{-\frac{1}{2}}
\, .
  \label{eqn:uw_def}
\end{split}
\end{equation}
Using the scalings~\eqref{eqn:Delta_lambda_scale} and~\eqref{eqn:g_scale}, one has the vertex function scale as follows:
\begin{equation}
    \mathcal{A}_{\phi NN \pi \pi} \sim \frac{4\pi}{m_N} \frac{1}{\sqrt{\mhi}} \frac{1}{f_\pi}\, . \label{eqn:phiNNpipi_scale}
\end{equation}

\begin{figure}[tb]
  \centering
  \includegraphics[scale=0.9]{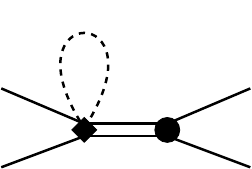}
  \caption{Irreducible $NN$ diagram built out of a $\phi NN \pi \pi$ vertex.}
  \label{fig:pionbubble}
\end{figure}

Combining then the scaling~\eqref{eqn:phiNNpipi_scale} with the standard counting rule for irreducible pion loops, we find that the $\phi NN \pi \pi$ vertex contributes to the two-nucleon force at {\NNLO}, illustrated in Fig.~\ref{fig:pionbubble}. However, the amplitude of this diagram vanishes, which can be seen by evaluating the isospin factors of $\mathcal{A}_{\phi NN \pi\pi}$ under exchange of the pion isospin index $a$ and $b$:
\begin{equation}
    \epsilon_{aac} = 0, \quad \text{and} \quad \delta_{ac} \mathcal{P}_a - \delta_{ac} \mathcal{P}_a = 0 \, .
\end{equation}

\begin{figure}
  \centering
  \includegraphics[scale=0.8]{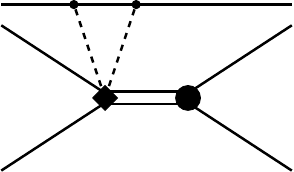}
  \hspace{35pt}
  \includegraphics[scale=0.8]{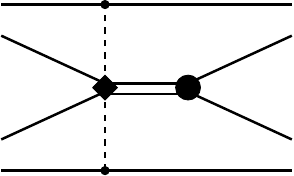}
  \caption{The Feynman diagrams contributing to three-body and four-body forces with the $\phi NN \pi \pi$ vertex.
}
  \label{fig_cont3b}
\end{figure}

As shown in Fig.~\ref{fig_cont3b}, the $\phi NN \pi \pi$ vertex~\eqref{eqn:AphiNNpipi} contributes to three-nucleon forces and four-nucleon forces. If we follow Weinberg's counting for long-range few-nucleon forces, these both enter at N$^4$LO, four orders higher than the LO two-nucleon forces. If one ever ventures to describe two-pion production by nucleon-nucleon collision, the $\phi NN \pi \pi$ will become relevant. These contributions are either quite high-order effects ($3N$ or $4N$ forces) or become sizable only at such high momenta ($NN \to NN \pi \pi$) where the applicability of chiral EFT is questionable at best. In conclusion, the enhancement due to the momentum-dependent dibaryon vertex does not give rise to particularly significant extraneous contributions to nuclear structures or reactions.

\section{Summary and conclusions\label{sec:Summary}}

We have considered a new formulation of chiral EFT to improve the slow convergence of the $\cs{1}{0}$ phase shifts. The key idea is to replace the energy-dependent dibaryon potential proposed in Ref.~\cite{Long:2013cya} with a momentum-dependent formulation that it is more amenable to many-body methods used for nuclear-structure calculations.

The main ingredient is a transition form factor, from $\cs{1}{0}$ $NN$ states to the spin-0 and isospin-1 dibaryon:
\begin{equation}
    \mathcal{A}_{\phi NN} = -\frac{\sqrt{4\pi} g}{\sqrt{p^2/\Delta + m_N}} \mathcal{P}_m
    \label{eqn:phiNN_vertex_conclusion}
\end{equation}
with two parameters $g$ and $\Delta$ characterizing, respectively, the strength and momentum range of the transition. The vertex function~\eqref{eqn:phiNN_vertex_conclusion} is interpreted as the sum of an infinite sequence of $\phi NN$ derivative-coupling operators, correlated by $g$ and $\Delta$, which in turn correspond to two independent low-energy mass scales, as shown in Sec.~\ref{sec:sprpot}. With $g$ and $\Delta$ varying, the resulting momentum-dependent potential traces out a two-dimensional surface in the EFT parameter space, in contrast to the one-dimensional line for the more conventional $\cs{1}{0}$ LO chiral forces.

This phenomenologically inspired potential must demonstrate its consistency with the chiral Lagrangian. For instance, the $\phi NN$ vertex can be radiatively corrected by the pion and could contribute non-trivial force to $\cs{1}{0}$ at {\NNLO} [see Fig.~\ref{fig:radpi_diagram} (a)]. When one of the incoming nucleons is connected to one of the outgoing nucleons by a radiation pion, triplet-channel amplitudes receive corrections as well [see Fig.~\ref{fig:radpi_diagram} (b)]. We examined these radiative corrections and found that they turn out to modify the long-range chiral forces only modestly.  In particular, they leave the already satisfactory convergence in the triplet channels intact.

We investigated further along this line, studying the case where a third nucleon is present. More specifically, the triton binding energy was calculated up to NLO using a power-counting scheme with partially perturbative pions~\cite{Long:2011xw,Wu:2018lai}.

Since the $\phi NN$ transition form factor is interpreted as an infinite series of nucleonic derivative couplings, it is accompanied by a $\phi NN \pi \pi$ vertex in order to satisfy chiral symmetry. We worked out the expression of this $\phi NN \pi \pi$ vertex function and discussed its phenomenological impacts, concluding that it only modifies chiral nuclear forces at quite high orders.

Because of our practical motivation of nuclear-structure studies, we did not touch the issue of chiral extrapolation, analytical continuation of observables in the light quark mass $m_{u, d}$. The chiral symmetry-breaking parts of $\cs{1}{0}$ contact terms have been argued to defy NDA based on RG considerations~\cite{Kaplan:1998we}. We conjecture here that the UV-suppressing dibaryon vertex offers an intriguing opportunity to reduce the divergences proportional to $m_\pi^2$, and then this might change the degree to which RG invariance modifies the NDA-based power counting. However, it is unclear how this can be implemented in a model-independent fashion.

In closing we note that likely further versions of a multi-parameter LO in the $\cs{1}{0}$ channel can be constructed with yet another formulation.
What was done in this paper for the momentum-dependent $\cs{1}{0}$ dibaryon potential provides guidance for inspecting whether the specific choice of interaction can be incorporated into chiral EFT.

\begin{acknowledgements}
We thank Bira van Kolck for valuable comments on the manuscript.
BL and SK thank the Institute for Nuclear Theory at the University of Washington and the organizers of the program ``Nuclear Structure at the Crossroads'' for hospitality while part of this work was carried out. This work was supported by the National Natural Science Foundation of China (NSFC) under Grant Nos. 11775148 and 11735003, the Fundamental Research Funds for the Central Universities, and by the National Science Foundation under Grant No.~PHY--2044632.
This material is based upon work supported by the U.S.\ Department of Energy,
Office of Science, Office of Nuclear Physics, under the FRIB Theory Alliance,
Award No. DE-SC0013617.
Computational resources for parts of this work have been provided by the Jülich Supercomputing Center.
\end{acknowledgements}

\appendix

\section{Vertex function of \texorpdfstring{$NN \pi \pi \to \phi$}{NNpipiphi}\label{sec:Appdx_Feyn}}

In order to satisfy chiral symmetry, there must be pion-absorption vertices entering along with $\mathcal{A}_{NN\phi}$ defined in Eq.~\eqref{eqn:phiNN_vertex}. In this appendix, we work out the vertex function for the transition $NN \pi \pi \to \phi$, discussed in Sec.~\ref{sec:chiral}.

A nonlinearly realized chiral transformation acts on the nucleon field like an isospin rotation with an angle depending on the local pion field. Let us focus on the axial sector $SU(2)_A$ of chiral symmetry, parametrized by $\bm{\theta}^A$ with stereographic coordinates:
\begin{equation}
  \begin{split}
    \delta_A N &= - i \left(\bm{\theta}^A \times \frac{\bm{\pi}}{2f_\pi}\right) \cdot \frac{\bm{\tau}}{2} N \,, \\
    \delta_A \bm{\pi} &= f_\pi \bm{\theta}^A\left(1 - \frac{\bm{\pi}^2}{4f_\pi^2}\right) + \left(\frac{\bm{\theta}^A\cdot\bm{\pi}}{2f_\pi} \right) \bm{\pi} \,, \\
     (\delta_A \bm{\phi})_c &= \left[\left(\bm{\theta}^A \times \frac{\bm{\pi}}{2f_\pi}\right) \times \bm{\phi}\right]_c \,.
  \end{split}
\end{equation}
We first seek the chiral-connection operator that has two pion fields, the fewest number of pion fields required by chiral symmetry, and then study its vertex function in momentum space.

As an illustration of our approach, we first derive the chiral-connection terms for the following operator:
\begin{equation}
  \sum_{m} \phi_m^\dagger\left[N^T \mathcal{P}_m \overleftrightarrow{\nabla}^2 N \right] \,. \label{phi_nucl_2ed}
\end{equation}
Unless noted otherwise, derivatives here and in the following act only on the fields within brackets. $\phi_m^\dagger$ in the above equation, for instance, is not acted upon. The most straightforward way to account for chiral symmetry is to replace the derivative $\overleftrightarrow{\nabla}$ with the chiral-covariant derivative $\overleftrightarrow{\mathscr{D}}$ defined in Eq.~\eqref{eq:covdervt}. Expanding in powers of $\bm{\pi}$, we can read off the corresponding chiral-connection terms up to $\mathcal{O}(\bm{\pi}^2/f_\pi^2)$:
\begin{equation}
  \begin{split}
    \sum_{b\, c\, d\, m} \frac{i\epsilon_{bcd}}{4f_\pi^2}  \phi_m^\dagger \sum_i \biggl\{ & \Bigl[ 2\left(\nabla_i N \right)^T\tau_d^T \mathcal{P}_m \pi_b (\nabla_i \pi_c)N + N^T\tau_d^T \mathcal{P}_m \pi_b \left(\nabla^2 \pi_c\right)N  \\
    & - 2N^T\tau_d^T \mathcal{P}_m \pi_b (\nabla_i \pi_c)(\nabla_i N) \Bigr] + \Bigl[2N^T \mathcal{P}_m\tau_d \pi_b (\nabla_i\pi_c)(\nabla_i N) \\
     & + N^T \mathcal{P}_m\tau_d \pi_b \left(\nabla^2 \pi_c \right)N - 2(\nabla_iN)^T \mathcal{P}_m\tau_d \pi_b (\nabla_i\pi_c)N \Bigr] \biggr\}\,. \\
  \end{split}
\end{equation}

However, the expansion becomes cumbersome if we substitute the chiral-covariant derivatives in the following operator, which is part of the Lagrangian~\eqref{eqn:lagrangian}, when $n$ is large:
\begin{equation}
  \sum_{m} \phi_m^\dagger \left[ N^T \mathcal{P}_m(-i\overleftrightarrow{\nabla})^{2n} N \right]
  \, . \label{eqn:deriv2n}
\end{equation}
We therefore try a slightly different approach and consider the axial transform $\delta_A$ of the operator in Eq.~\eqref{phi_nucl_2ed}:
\begin{equation}
  \begin{split}
   \sum_{b\, c\, d\, m} \phi_{m}^{\dagger} & \left[N^{T} \mathcal{P}_{m}\left(\overleftarrow{\nabla}-\overrightarrow{\nabla}\right)^{2} \left(-i \epsilon_{b c d} \theta_{b} \frac{\pi_{c}}{4 f_{\pi}} \tau_{d} N \right) \right. \\
    & \left. + N^{T} \left(-i \epsilon_{b c d} \theta_{b} \frac{\pi_{c}}{4 f_{\pi}}  \right) \tau_{d}^T \mathcal{P}_{m}\left(\overleftarrow{\nabla}-\overrightarrow{\nabla}\right)^{2} N \right] \,. \label{axial_nucl}
\end{split}
\end{equation}
Noticing that to the lowest order in $\bm{\pi}^2/f_\pi^2$ we have
\begin{equation}
    \delta_A \bm{\pi} = f_\pi \bm{\theta}^A + \mathcal{O}(\bm{\pi}^2/f_\pi^2)\, ,
\end{equation}
we tentatively propose the following operator to cancel the preceding chiral-symmetry violation~\eqref{axial_nucl}:
\begin{equation}
  \begin{split}
     \sum_{b\, c\, d\, m} \frac{i}{4 f_{\pi}^{2}} \epsilon_{b c d} \pi_{b} \phi_{m}^{\dagger} & \left\{ \left[N^{T} \mathcal{P}_m \tau_{d}\left(\overleftarrow{\nabla}-\overrightarrow{\nabla}\right)^{2} \pi_{c} N - N^T \mathcal{P}_m \tau_d \overleftarrow{\nabla}^{2} \pi_{c} N \right] \right. \\
    & \left. + \left[N^{T} \pi_{c} \tau_{d}^T \mathcal{P}_m \left(\overleftarrow{\nabla}-\overrightarrow{\nabla}\right)^{2} N - N^T \pi_{c} \tau_d^T \mathcal{P}_m \overrightarrow{\nabla}^{2} N \right] \right\} \, . \label{conct_2ed}
  \end{split}
\end{equation}
It is straightforward to verify that the combination of the operators in Eqs.~\eqref{phi_nucl_2ed} and~\eqref{conct_2ed} is chiral invariant up to $\mathcal{O}(\bm{\pi}^2/f_\pi^2)$.

Coming back to the more general case, we expand Eq.~\eqref{eqn:deriv2n} as
\begin{multline}
 \sum_m \phi_{m}^{\dagger}\left[N^T \mathcal{P}_{m}({-}i\overleftrightarrow{\nabla})^{2 n} N\right] \\
  = \sum_m \sum_{k+l+q=n} C(k, l, q) \phi_{m}^{\dagger} \left[N^T \mathcal{P}_m \left(\frac{\overleftarrow{\nabla}}{i} \right)^{2k} \left({-} \frac{\overleftarrow{\nabla}}{i} \cdot \frac{\overrightarrow{\nabla}}{i} \right)^l \left(\frac{\overrightarrow{\nabla}}{i} \right)^{2q} N\right] \,,
 \label{phi_nucl_nth}
\end{multline}
where $C(k, l, q)$ are binomial coefficients,
\begin{equation}
  \left(x+y \right)^{2n} = \sum_{k+l+q=n} C(k, l, q) x^{2k} (xy)^l y^{2q} \, ,
\end{equation}
satisfying
\begin{equation}
  C(k, l, q) = C(q, l, k) \,. \label{eqn:bin_rela}
\end{equation}
The chiral-connection operator for Eq.~\eqref{phi_nucl_nth}, in analogy with Eq.~\eqref{conct_2ed}, is
\begin{equation}
  \sum_{k+l+q=n} C(k, l, q)\left[\mathcal{O}_B(k,l,q) + \mathcal{O}_C(k,l,q) \right] \,, \label{conct_nth}
\end{equation}
where
\begin{equation}
\begin{split}
    \mathcal{O}_B(k,l,q)\equiv \sum_{b\, c\, d\, m} \frac{i}{4 f_{\pi}^{2}} \epsilon_{bcd} \pi_{b} \phi_m^\dagger \left[N^T \mathcal{P}_m  \tau_d \left(\frac{\overleftarrow{\nabla}}{i}\right)^{2k} \left( - \frac{\overleftarrow{\nabla}}{i} \cdot \frac{\overrightarrow{\nabla}}{i} \right)^l \left(\frac{\overrightarrow{\nabla}}{i}\right)^{2q} \pi_c N \right] \,, \\
    \mathcal{O}_C(k,l,q)\equiv \sum_{b\, c\, d\, m} \frac{i}{4 f_{\pi}^{2}} \epsilon_{bcd} \pi_{b} \phi_m^\dagger \left[ N^T \pi_c \tau_d^T \mathcal{P}_m \left(\frac{\overleftarrow{\nabla}}{i}\right)^{2k} \left( - \frac{\overleftarrow{\nabla}}{i} \cdot \frac{\overrightarrow{\nabla}}{i} \right)^l \left(\frac{\overrightarrow{\nabla}}{i}\right)^{2q} N \right] \, .
\end{split}
\label{eq:chiralcon_oboc}
\end{equation}
For $\mathcal{O}_B(k, l, q)$, either $l$ or $q$ must be positive so that there is at least one $\overrightarrow{\nabla}$ acting on $\pi_c N$. Similarly, either $l$ or $k$ must be positive for $\mathcal{O}_C(k,l,q)$. Transposing the operator inside the square brackets, say, of $\mathcal{O}_B$ and using $\mathcal{P}_a^T=-\mathcal{P}_a$, we realize
\begin{equation}
  \mathcal{O}_C(k,l,q) = \mathcal{O}_B(q,l,k) \,.
\end{equation}
Combining this with Eq.~\eqref{eqn:bin_rela}, one concludes that Eq.~\eqref{conct_nth} reduces to
\begin{equation}
  2\sum_{k+l+q=n} C(k,l,q) \mathcal{O}_B(k,l,q) \,. \label{eq:chiral_con_opt}
\end{equation}

We then consider the following matrix element of $\mathcal{O}_B(k,l,q)$ between the free-particle states whose quantum numbers are specified in Fig.~\ref{fig:phiNNpipi}:
\begin{multline}
    \int \mathrm{d}^{4}x \langle \phi_{c}, p_{3} | \mathcal{O}_{B}(k, l, q; x) | k_{1},a; k_{2},b; p_{1}, \alpha_{1}; p_{2}, \alpha_{2} \rangle \\
    = \frac{i}{4 f_{\pi}^{2}} \sum_{b^\prime\, c^\prime\, d^\prime\, m} \epsilon_{b^\prime c ^\prime d^\prime} \int \mathrm{d}^{4}x \langle \phi_{c}, p_{3} |
    \pi_{b^\prime}\phi_m^\dagger \left[N^T \mathcal{P}_m  \tau_{d^\prime} \left(\frac{\overleftarrow{\nabla}}{i}\right)^{2k} \left(- \frac{\overleftarrow{\nabla}}{i} \cdot \frac{\overrightarrow{\nabla}}{i} \right)^l \left(\frac{\overrightarrow{\nabla}}{i}\right)^{2q} (\pi_{c^\prime} N) \right]_x \\
    \times | k_{1},a; k_{2},b; p_{1}, \alpha_{1}; p_{2}, \alpha_{2} \rangle \,.
\end{multline}
Here $a,b,c$ are the isospin indexes of the external pions and intermediate dibaryon. After counting all contractions and stripping the $\delta$ function that enforces momentum conservation we obtain
\begin{equation}
\begin{split}
    -\frac{i}{4f_{\pi}^{2}} & \Biggl\{ \left[\frac{1}{\sqrt{8}} \epsilon_{abc} \sigma_{2} \tau_{2} + i \left(\delta_{bc}\mathcal{P}_a - \delta_{ac}\mathcal{P}_b \right) \right]_{\alpha_2,\alpha_1} \\
    & \times \vec{p}_{2}^{\,2 k} \left\{\left[-\vec{p}_{2} \cdot\left(\vec{p}_{1}+\vec{k}_{1}\right)\right]^{l}\left(\vec{p}_{1}+\vec{k}_{1}\right)^{2 q} -\left[-\vec{p}_{2} \cdot\left(\vec{p}_{1}+\vec{k}_{2}\right)\right]^{l}\left(\vec{p}_{1}+\vec{k}_{2}\right)^{2 q}\right\} \\
    & + \left[-\frac{1}{\sqrt{8}} \epsilon_{abc} \sigma_{2} \tau_{2} + i \left(\delta_{bc}\mathcal{P}_a - \delta_{ac}\mathcal{P}_b \right) \right]_{\alpha_2,\alpha_1} \\
    & \times \vec{p}_{1}^{\,2 k} \left\{\left[ - \vec{p}_{1} \cdot\left(\vec{p}_{2}+\vec{k}_{1}\right)\right]^{l}\left(\vec{p}_{2}+\vec{k}_{1}\right)^{2 q} -\left[ - \vec{p}_{1} \cdot\left(\vec{p}_{2}+\vec{k}_{2}\right)\right]^{l}\left(\vec{p}_{2}+\vec{k}_{2}\right)^{2 q}\right\} \Biggr\} \,.
\end{split}
\label{eq:feyn_oboc}
\end{equation}
The dibaryon momentum $\vec{p}_3$ does not appear in the above equation because the derivatives in Eq.~\eqref{eq:chiralcon_oboc} act only on the fields inside the brackets.

To obtain the $\phi NN\pi\pi$ vertex function $\mathcal{A}_{\phi NN \pi \pi}$~\eqref{eqn:AphiNNpipi}, we need to carry out two summations of the above matrix elements. The first is over all the combinations of $(k, l, q)$ prescribed by Eq.~\eqref{eq:chiral_con_opt}, using Eq.~\eqref{eqn:bin_rela}. The second is given by summing $n$ from $0$ to $\infty$, as defined in the Lagrangian~\eqref{eqn:lagrangian}, using the identity
\begin{equation}
    \sum_{n=0}^{\infty} g_{2n} t^{2n} = g \sum_{n=0}^{\infty} \binom{-\frac{1}{2}}{n} t^{2n} = g\,W(t)
\end{equation}
with
\begin{equation}
    W(t) \; \equiv \; \left(1 + t^2\right)^{-\frac{1}{2}} \,.
\end{equation}
We can apply these two summations to any individual term in the expression~\eqref{eq:feyn_oboc}. Take the following term as an example. The first summation is given by
\begin{equation}
    \sum_{k+l+q=n} C(k,l,q)\, \vec{p}_{2}^{\,2 k} \left[-\vec{p}_{2} \cdot\left(\vec{p}_{1}+\vec{k}_{1}\right)\right]^{l}\left(\vec{p}_{1}+\vec{k}_{1}\right)^{2 q} = \left(\vec{p}_1 - \vec{p}_2 + \vec{k}_1 \right)^{2n} \, ,
\end{equation}
followed by the second summation:
\begin{equation}
    \sum_{n=0}^{\infty} g_{2n} \left[ \frac{(2\vec{p} + \vec{k}_1 )^2}{4m_N\Delta} \right]^n = g\, W\left(\frac{\left|p + \vec{k}_1/2 \right|}{\sqrt{m_N \Delta}}\right) \,.
\end{equation}
Adding up the contributions from every term in Eq.~\eqref{eq:feyn_oboc}, we eventually arrive at Eq.\eqref{eqn:AphiNNpipi}.

\bibliography{Separable.bib}

\end{document}